\documentclass[usenatbib]{mn2e}

\usepackage{amsmath}
\usepackage{graphics}
\usepackage{graphicx}
\usepackage{natbib}
\newcommand{\snia}{SN~Ia}
\newcommand{\sneia}{SNe~Ia}

\newcommand{\kms}{km\,s$^{-1}$}

\def\lesssim{\mathrel{\hbox{\rlap{\hbox{\lower3pt\hbox{$\sim$}}}\hbox{\raise2pt\hbox{$<$}}}}}
\def\gtrsim{\mathrel{\hbox{\rlap{\hbox{\lower3pt\hbox{$\sim$}}}\hbox{\raise2pt\hbox{$>$}}}}}

\newcommand{\logm}{$\log(M_*/M_\odot)$}

\title[SN Ia Ages Over Cosmic Time]{Ages of Type Ia Supernovae Over Cosmic Time}

\author[Childress, Wolf, \& Zahid]{
Michael~J.~Childress$^{1,2}$\thanks{E-mail:michael.childress@anu.edu.au},
Christian~Wolf$^{1,2}$,
H.~Jabran~Zahid$^{3,3}$ \\
$^{1}$  Research School of Astronomy and Astrophysics, 
Australian National University, 
Canberra, ACT 2611, Australia.\\
$^{2}$  ARC Centre of Excellence for All-sky Astrophysics (CAASTRO), 
Australian National University, 
Canberra, ACT 2611, Australia.\\
$^{3}$  Institute for Astronomy, University of Hawaii, 2680 Woodlawn Drive, Honolulu, HI 96822, USA.\\
$^{4}$  Smithsonian Astrophysical Observatory, Harvard-Smithsonian Center for Astrophysics, 60 Garden St., Cambridge, MA 02138, USA.
}

\begin{document}
\maketitle

%%%%%%%%%%%%%%%%%%%%%%%%%%%%%%%%%%%%%%%%%%%%%%%%%%%%%%%%%%%%%%%%%%%%%%%%%%%%%
\begin{abstract}
%------------------
We derive empirical models for galaxy mass assembly histories, and convolve these with theoretical delay time distribution (DTD) models for Type Ia supernovae (\sneia) to derive the distribution of progenitor ages for all \sneia\ occurring at a given epoch of cosmic time. 
% SF = consistently young, passive = diversely old
In actively star-forming galaxies, the progression of the star formation rate is shallower than a $t^{-1}$ \snia\ DTD, so mean \snia\ ages peak at the DTD peak in all star-forming galaxies at all epochs of cosmic history. In passive galaxies which have ceased star formation through some quenching process, the \snia\ age distribution peaks at the quenching epoch, which in passive galaxies evolves in redshift to track the past epoch of major star formation. 
% data/model comparison
Our models reproduce the \snia\ rate evolution in redshift, the relationship between \snia\ stretch and host mass, and the distribution of \snia\ host masses in a manner qualitatively consistent with observations.
% comment on HR trend!
Our model naturally predicts that low-mass galaxies tend to be actively star-forming while massive galaxies are generally passive, consistent with observations of galaxy ``downsizing''. Consequently, the mean ages of \sneia\ undergo a sharp transition from young ages at low host mass to old ages at high host mass, qualitatively similar to the transition of mean \snia\ Hubble residuals with host mass. The age discrepancy evolves with redshift in a manner currently not accounted for in \snia\ cosmology analyses. 
% use only SNe Ia in active hosts!
We thus suggest that \sneia\ selected only from actively star-forming galaxies will yield the most cosmologically uniform sample, due to the homogeneity of young \snia\ progenitor ages at all cosmological epochs.
\end{abstract}


%%%%%%%%%%%%%%%%%%%%%%%%%%%%%%%%%%%%%%%%%%%%%%%%%%%%%%%%%%%%%%%%%%%%%%%%%%%%%
\section{Introduction}
\label{sec:intro}
%---------------------------
Type Ia supernovae (\sneia) are believed to result from the thermonuclear disruption of a carbon-oxygen white dwarf (CO-WD) which reaches some instability condition through interaction with a binary companion \citep{hf60}. One possible scenario is the single-degenerate scenario \citep[SD;][]{whelan73, nomoto82}, where a CO-WD accretes material from a non-degenerate main sequence (MS) star or red giant (RG) companion until reaching the critical Chandrasekhar mass. Alternatively, in the double-degenerate (DD) scenario, two CO-WDs coalesce after losing angular momentum to gravitational waves \citep{tutukov76, tutukov79, iben84, webbink84}.  The binary evolution in these scenarios operates over a wide range of timescales from a few hundred Myr to a Hubble time. This manifests as a different instantaneous SN rate as a function of progenitor age, referred to as the delay time distribution \citep[DTD; for a review see][]{mm12}.  
%---------
% main objective of the paper
Though it is never possible to pinpoint the exact progenitor age giving rise to a particular supernova, we will demonstrate how it is possible to derive the distribution of progenitor ages for a large sample of supernovae by combining the DTD with the star formation history of the parent stellar population of the SNe.

% rates studies and DTD measurements
The existence of \sneia\ in passive host galaxies led to early speculation that \sneia\ could arise primarily from old stellar populations. However, \snia\ rates studies over the last decade suggest that the \snia\ rate in a galaxy is best parametrized by both the total stellar mass in the galaxy and its star formation rate \citep{mannucci05, scan05, sullivan06, mannucci06, aubourg08, li11, smith12}. This led to the ``two-component'' (or ``A+B'') model description wherein \sneia\ can arise from both old (or ``tardy'') and young (or ``prompt'') progenitors. Further measurements of \snia\ rates show that the \snia\ DTD appears consistent with a $t^{-1}$ power law \citep{totani08, maoz11, graur13}, such that the prompt \snia\ progenitors have a much higher representation than the old tardy progenitors \citep{forster06}.

% age distribution!
% - here need SFH models
The integrated rate of \sneia\ in a particular galaxy is the convolution of the \snia\ DTD with the star formation history (SFH) of the host galaxy \citep{yl00}. The integrand of that convolution (i.e. the product of the DTD and the galaxy's stellar age distribution) represents the likelihood function for the age of a given \snia\ in that galaxy. Equivalently, this represents the progenitor age distribution for a large sample of \sneia\ drawn from stellar populations with that mean SFH. To fully exploit the power of this progenitor age distribution requires a knowledge of galaxy SFHs, which are observationally challenging to constrain. Instead, we will show how knowledge of galaxy star formation rates (SFR) throughout cosmic time can constrain galaxy stellar mass assembly histories. These in turn can be coupled to an observationally-motivated \snia\ DTD to produce \snia\ progenitor age distributions at all galaxy mass scales and all redshifts.

% age vs. mass and Hubble Residual bias
% - galaxy mass as primary variable!
In this work we focus on how \snia\ ages depend on the stellar mass of their host galaxies, from low-mass dwarf galaxies (\logm\ $\sim8$) to the most massive (\logm\ $\sim12$) ellipticals. Throughout this work, we refer to this as the ``galaxy mass sequence''.
% - Hubble Residuals!!
The progression of \snia\ progenitor ages along the galaxy mass sequence, and its evolution with redshift, have critical implications for the use of \sneia\ as cosmological distance indicators. Observations of \sneia\ led to the discovery of the accelerating expansion of the Universe \citep{riess98, perlmutter99} and continue to constrain cosmological parameters \citep[e.g.,][]{sullivan11a, suzuki12, rest13, betoule14}. While systematic uncertainties had for a time constituted the dominant fraction of the \snia\ cosmology error budget \citep{scolnic13, scolnic14a, betoule13, mosher14}, significant effort has been expended to reduce systematics so that statistical error now dominates the error budget \citep{betoule14}.  This statistical error arises from the ``intrinsic scatter'' in corrected \snia\ luminosities, making the search for possible astrophysical drivers of this scatter of paramount importance.

More concerning is the recent discovery of a bias in the standardized \snia\ luminosities with the properties of their host galaxies. These luminosities, quantified by deviations from the best-fit cosmology on the Hubble Diagram (``Hubble residuals''), were found to depend on the mass (and/or metallicity) of their host galaxies \citep{sullivan10, kelly10, lampeitl10, gupta11, dandrea11, konishi11, galbany12, hayden13, johansson13, childress13b}.  \citet{childress13b} inspect the trend of Hubble residuals along the galaxy mass sequence and find age to be the physical property most consistent with the observed trend. \citet{johansson13} reach a similar conclusion. Furthermore, based on the analysis of SF intensity at \snia\ locations, \citet[][hereafter R13]{rigault13} find this host bias is most likely driven by fast declining \sneia\ in passive regions within their host galaxies \citep[though these \sneia\ could cease to be problematic with the use of a properly trained \snia\ light curve fitter, see e.g.][]{kim13, kim14}. These results indicate that progenitor age could impart a critical bias on the measurement of cosmological parameters with \sneia. 
%Here we will derive the evolving \snia\ progenitor age distribution and examine its cosmological implications.

% paper outline
In Section~\ref{sec:mass_buildup} we describe our empirical model for the buildup of stellar mass in galaxies over cosmic time and in Section \ref{sec:age_dists} we use this model to examine the progenitor age distribution of \sneia\ from $z=2$ to the present epoch. The mean ages of \sneia\ as a function of their host galaxy stellar mass is presented in Section~\ref{sec:age_vs_mass}. We then show that the trend of \snia\ age with host mass is robust against uncertainties in the stellar mass assembly history of galaxies (Section~\ref{sec:sfh_uncertainties}) or the exact form of the \snia\ DTD (Section~\ref{sec:dtd_uncertainties}). We summarize and conclude in Section~\ref{sec:conclusions}. For the required calculation of lookback time as a function of redshift, we employ the standard ``concordance'' $\Lambda$CDM cosmology with $\Omega_M=0.30$, $H_0=70$~\kms~Mpc$^{-1}$, and $\Omega_\Lambda=0.70$.

%%%%%%%%%%%%%%%%%%%%%%%%%%%%%%%%%%%%%%%%%%%%%%%%%%%%%%%%%%%%%%%%%%%%%%%%%%%%%
\section{The Stellar Mass Dependence of Galaxy Star Formation Histories}
\label{sec:mass_buildup}
%---------------------------
% summary
Understanding the mass assembly and star formation history of galaxies over cosmic time is a rich and vigorous field of study \citep[for a recent review see][]{madau14}. In this Section we utilize observational constraints on star formation (SF), mass recycling, and SF quenching to forward model the buildup of stellar mass in galaxies over cosmic time. Specifically, we focus on the mean SFH of galaxies as a function of their stellar mass at all redshifts. The quantitative parametrizations of all galaxy scaling relations employed in our models are presented in full in Appendix~\ref{app:equations}. This Section focuses on describing the qualitative features of our model, while the pertinent equations from Appendix~\ref{app:equations} are referred to parenthetically for reference. We show later (Section~\ref{sec:sfh_uncertainties}) that the main results of this work are highly insensitive to the specific parametrizations employed for galaxy stellar mass assembly.

\begin{figure*}
\begin{center}
\includegraphics[width=0.90\textwidth]{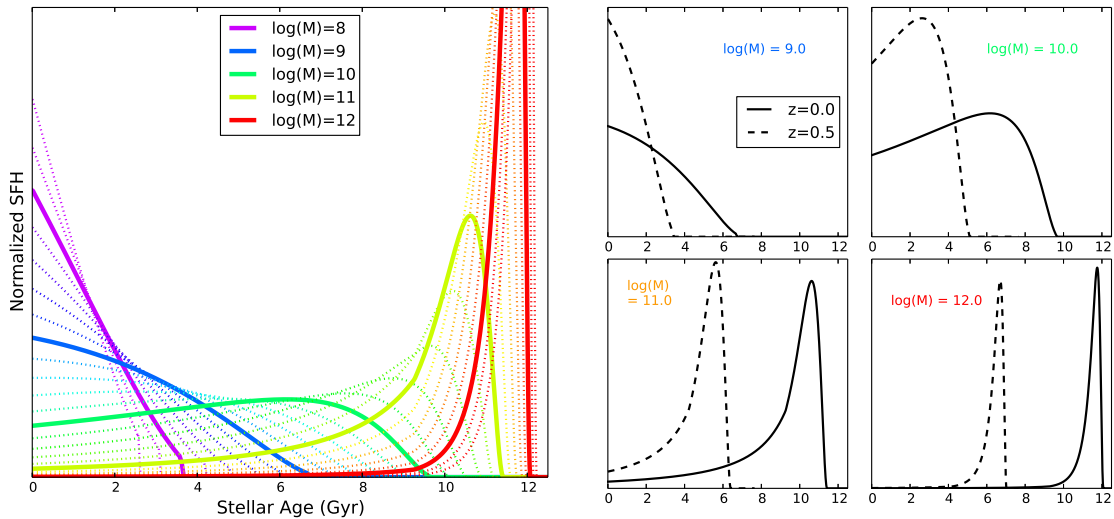}
\caption{{\em Left:} Final mean SFHs along the galaxy mass sequence for our models, in steps of 0.1~dex from \logm$=7.5$ to \logm$=12.5$.  Integer log stellar mass values are highlighted as thick solid curves.  {\em Right:} Mean SFHs at the current epoch ($z=0$) and at the same mass scale 5~Gyr in the past ($z=0.5$) for select final stellar mass values.}
\label{fig:example_sfhs}
\end{center}
\end{figure*}

%---------------------------
% sfr vs. mass
In the local Universe, the star formation rate is strongly correlated with the stellar mass \citep{salim07, elbaz07} and similar trends are seen out to $z\sim2$ \citep[e.g.,][Z12 hereafter]{noeske07, pannella09, karim11, whitaker12, kashino13, zahid12b}. The slope and scatter of the relation between stellar mass and SFR is not strongly dependent on redshift \citep[e.g., Z12;][]{noeske07, whitaker12}. These observations provide strong constraints for the stellar mass buildup of galaxies \citep[e.g., Z12;][]{leitner12}. For our stellar mass assembly tracks, we use the Z12 parametrization of the relationship between SFR, stellar mass, and redshift (hereafter the ``SMz relation'').

%---------------------------
% quenching
Eventually galaxies use up (or perhaps lose) their gas which could be converted into stars, and star formation effectively ceases (or ``quenches''). The mass functions of ``active'' star-forming galaxies and ``passive'' galaxies with little or no star formation (typically separated using color-magnitude diagrams) reach equality at $z\sim0$ at a mass scale of \logm$\sim10.0-10.4$. This ``quenching mass'' has been measured in the local universe by GAMA \citep{baldry12} and SDSS \citep{bell03, baldry04}; at intermediate ($z \lesssim 2$) redshifts by deep surveys such as COMBO-17 \citep{borch06}, DEEP2 \citep{bundy06}, and NEWFIRM \citep{brammer11}; and more recently at very high redshifts ($2 \lesssim z \lesssim 4$) with the COSMOS/UltraVISTA survey \citep{muzzin13}. The dependence of this quenching mass with redshift has been fit by \citet{muzzin13}, and we use this redshift dependence (Equation~\ref{eq:mq_vs_z}) in our baseline mass assembly models.

% quenching transition with mass
The quenching mass scale in Equation~\ref{eq:mq_vs_z} represents the \emph{median} quenching mass, when active and passive galaxies are equally represented. For our models we require a prescription for the passive galaxy fraction as a function of galaxy stellar mass. This can be explicitly calculated when the Schechter function parameters for active and passive galaxies are well-measured, but is challenging when these are poorly constrained. Using the blue and red galaxy mass functions at low redshift ($z=0.1$) where the active and passive Schechter function parameters are well measured from \citet{moustakas13}, we find the passive fraction is accurately represented by an error function centered at the median quenching mass and with a width of 1.5 dex in log stellar mass (Equation~\ref{eq:quenching_penalty}). To simplify the modeling process, we thus adopt this functional form for all redshifts with a redshift-dependent median quenching mass that follows Equation~\ref{eq:mq_vs_z}.

% does quenching really shut off SF?
We note, however, that the quenching mass scale derived from color-based separation of active and passive galaxies are likely to be slightly biased to lower values. This is due to the high-mass end of the star-forming main sequence having some overlap with (and thus contaminating) the red galaxy sequence due to the combination of its low sSFR and reddening by dust. Dust reddening is also higher in high-mass star-forming galaxies \citep[e.g.][]{jlee09, garn10, zahid13b, zahid13c}, hence a significant tail of the SF galaxy population is likely to be wrongly attributed to the quenched population \citep[e.g.][]{wolf09}. This effect was first noticed in galaxy clusters \citep{wolf05} but extends to field galaxy samples as well. As a requisite example, \citet{spitler14} used multi-band SEDs for $z\sim3-4$ galaxies from the ZFOURGE survey to cleanly separate dusty and truly passive galaxies, and found that dusty star-forming galaxies may consitute nearly half of red massive galaxies \citep[see also][]{straatman14}. Moreover, the intrinsic color distributions of active and quiescent galaxies have some overlap and thus cannot be separated without contamination using a strict color cut \citep{taylor14}. We will revisit these concerns and their (negligible) implications for our conclusions in Section~\ref{sec:sfh_uncertainties}.

%---------------------------
% mass recycling
The calculation of a galaxy's integrated SFH must also take into account the fraction of gas converted into stars that will be recycled back to the interstellar medium (ISM) by supernovae or stellar winds. This recycling fraction depends on time because lower mass stars progressively evolve off the main-sequence and also depends on the stellar initial mass function (IMF). \citet{lk11} explored these effects in detail, and found that the fraction of material returned to the ISM rises rapidly in the first few Gyr, and then levels off. They found that the fraction of stellar mass returned to the ISM from a single burst of SF could be well parametrized by a shifted exponential function (Equation~\ref{eq:mass_loss}). For our analytical galaxy SFHs, we utilize this functional form along with the parameters derived by \citet{lk11} for a \citet{chab03} IMF.

%---------------------------
%---------------------------
% final mass assembly
% - unquenched regime
The above prescriptions provide the critical pieces necessary to model the buildup of stellar mass in galaxies:
\begin{enumerate}
  \item The dependence of SFR on current stellar mass and its evolution in redshift (Equation~\ref{eq:smz}).
  \item The fraction of stellar mass returned to the ISM as a function of time (Equation~\ref{eq:mass_loss}).
  \item The stellar mass scale at which star formation ceases, and its dependence on redshift (Equation~\ref{eq:mq_vs_z}).
  \item The fraction of galaxies which have ceased SF as a function of stellar mass (Equation~\ref{eq:quenching_penalty}).
\end{enumerate}
With these we can construct galaxy SFHs by integrating a galaxy's SFR over cosmic time (Equation~\ref{eq:mass_buildup}). Such a procedure was explored in detail in Z12, and we adopt a similar but slightly modified technique for deriving analytical galaxy SFHs along the galaxy mass sequence.
% MASS BUILDUP!
In practice, we begin with a set of mass evolution tracks starting from a mass of $M_*=10^6M_\odot$ at some time of formation $t_F$ then follow the SMz relation to the present epoch $z=0$. We model tracks with various values of $t_F$ (in steps of 50~Myr from 1-10~Gyr, and steps of 25~Myr from 10-13~Gyr of lookback time), and forward model the mass buildup with time steps of 0.5~Myr.

% - quenching penalty!
To obtain the SFH for galaxies of a given stellar mass, we must account for the effects of SF quenching. For this, we impose a ``quenching penalty'' on the SF of our model galaxies at each time step by multiplying the SF from the SMz by the passive galaxy fraction at the current galaxy mass scale at the epoch of that time step (see Equation~\ref{eq:mass_buildup}). 
%This results in our models preserving the net total SF across all galaxies at that mass scale at that redshift.

% FINALLY, PRESENT SFHS FOR VARIOUS MASSES!!
The final SFHs from our models are shown in Figure~\ref{fig:example_sfhs} in steps of 0.2~dex from \logm$=7.6$ to \logm$=12.4$.  The right panels of this Figure show SFHs for several specific stellar mass values at both the current epoch ($z=0$) and 5~Gyr in the past ($z=0.5$).  For higher mass galaxies, the mean SFHs generally are confined to a small distribution of old ages, reflecting the rapid transition to fully quenched SF.  Lower mass galaxies tend to be almost completely unquenched, and generally show SFRs that increase toward the current epoch.

% commentary on final SFHs
The SFR-mass trend, effects of SF quenching, and their respective evolutions in redshift, combine to paint a picture of SFHs along the galaxy mass sequence. Massive galaxies above the quenching mass at a given epoch formed the majority of their stars 5-10~Gyr in the past and effectively ceased SF at high redshift ($z\gtrsim2$), while less massive galaxies (\logm$\lesssim10$) formed more of their stars recently and indeed are still actively forming stars. The quenching epoch for massive galaxies is at roughly 10~Gyr in the past for massive $z=0$ galaxies and at 5~Gyr in the past for massive $z=0.5$ galaxies. This will result in important consequences for the mean progenitor ages for \sneia\ along the galaxy mass sequence, as we explore below in Section~\ref{sec:age_vs_mass}.

%---------------------------
% comparison to cosmic SFH
Though our models by construction obey the SMz relation and quenching fraction at all redshifts, it is worthwhile to confirm that they paint a consistent picture for the global SFH of the Universe. We therefore calculate our prediction for the SFH of the Universe by summing our mean SFHs weighted by the total stellar mass in each logarithmic galaxy mass bin at z=0. This is simply a modified Schechter function, for which we use the \citet{moustakas13} data (which we find is well-fitted by a double Schechter function with parameters $\alpha_1=-1.42$, $\log M_1^*=11.12$, $\phi_1^*=1.39\times10^{-3}$, $\alpha_2=-0.45$, $\log M_2^*=10.66$, $\phi_2^*=6.57\times10^{-3}$).

\begin{figure}
\begin{center}
\includegraphics[width=0.45\textwidth]{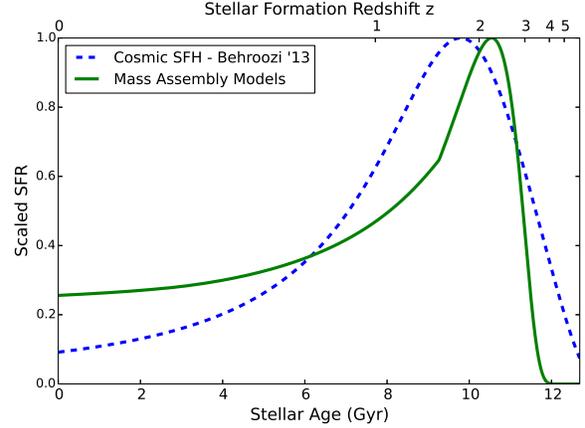}
\caption{Predictions for the volume-averaged star formation history of the Universe obtained from the Schechter function-weighted sum of our mass assembly models (solid green curve) compared to the cosmic SFH inferred from the compilation of SFR data presented in \citet{behroozi13}.}
\label{fig:models_vs_csfh}
\end{center}
\end{figure}

In Figure~\ref{fig:models_vs_csfh} we show our model's prediction for the cosmic SFH at $z=0$ compared to that derived from the compilation of observational data in \citet[][hereafter B13]{behroozi13}. As with the obesrved CSFH, our prediction for the CSFH shows a characteristic peak near $z\sim2$, and declines toward the present epoch.  Some discrepancy is evident here, and we show in Section~\ref{sec:sfh_uncertainties} that these differences can be reconciled with some adjustments to the galaxy mass assembly models.  It is worth noting that accurate prediction of the cosmic SFH is still a challenge for the galaxy evolution community, even with simulations that fully track dark matter structure growth \citep[e.g.,][]{millennium} with added presciptions for baryonic physics.  Our CSFH is produced solely from the mass-dependent galaxy SFHs from our models weighted by the observed number density of galaxies as a function of mass in the local Universe. That we obtain the level of qualitative agreement seen here is encouraging for the validity of our galaxy SFHs. Most importantly, in Section~\ref{sec:sfh_uncertainties} we show that any attempted adjustments to our mass assembly models cannot negate our conclusions regarding \snia\ ages and their relationship with host galaxy mass throughout cosmic time.

%Our cosmic SFH is qualitatively similar to the observed one, in that it rises sharply to its peak close to redshift $z=2$, then declines more slowly to the present epoch. Though there are still quantitative differences (indeed, the distribution of all stars formed throughout the history of the Universe amongst present day galaxies is an ongoing field of study), we consider the qualitative similarities sufficient to allow examination of \snia\ age distributions which form the central topic of this paper. We revisit possible alterations to the mass assembly details in Section~\ref{sec:sfh_uncertainties}.

%%%%%%%%%%%%%%%%%%%%%%%%%%%%%%%%%%%%%%%%%%%%%%%%%%%%%%%%%%%%%%%%%%%%%%%%%%%%%
\section{The Distribution of \snia\ Ages Over Cosmic Time}
\label{sec:age_dists}
%---------------------------
In this section we use our empirical galaxy SFHs from Section~\ref{sec:mass_buildup} to infer the \snia\ progenitor age distribution as a function of host galaxy mass and redshift. We begin by inspecting the \snia\ age distribution in individual galaxies in Section~\ref{subsec:snia_age_dist}. In Section~\ref{subsec:age_bimodality} we show how \snia\ progenitor age distributions across the galaxy mass sequence manifest a sharp bimodality in the global \snia\ progenitor age distribution. Finally in Section~\ref{subsec:age_dist_vs_z} we inspect how the \snia\ progenitor age distribution evolves in redshift.

%------------------------------------------------------
\subsection{\snia\ Age Distribution for a Single Galaxy}
\label{subsec:snia_age_dist}
%------------------------------------------------------
The rate of \sneia\ in a galaxy at a specific epoch $t_0$ is the convolution of the \snia\ DTD $\phi(\tau)$ with the galaxy SFH $\psi(\tau)$:
\begin{equation}
  R(t_0) = \int_0^\infty \phi(\tau)\,\psi(t_0-\tau)\,d\tau
\label{eq:integrated_snia_rate}
\end{equation}
While Equation~\ref{eq:integrated_snia_rate} gives the total integrated \snia\ rate in a given galaxy, the integrand is in fact the probability distribution for the age of the progenitor system for any \snia\ occurring in that galaxy:
\begin{equation}
  P(\tau; t_0) = \phi(\tau)\,\psi(t_0-\tau).
\label{eq:snia_age_pdf}
\end{equation}
When $\tau > t_0$, $\psi = 0$. Thus $P(\tau; t_0)$ is the likelihood at epoch $t_0$ of the SN arising from a progenitor system of age $\tau$, and $\psi(t_0-\tau)$ is effectively the age distribution of all stars in the parent galaxy (if one corrects for the stellar mass lost over time). The quantity $P(\tau; t_0)$ also represents the \emph{age distribution} for a large sample of \sneia\ arising from the same galaxy, or equivalently from many galaxies with the same SFH. 
% additive nature implies that the mean SFH is sufficient to give us 
% the correct age distribution!
Thus, the mean SFHs calculated in Section~\ref{sec:mass_buildup} are appropriate for calculating the age distribution for many \sneia\ occurring in host galaxies with the same stellar mass.

% DTD functional form!
In order to avoid sharp features in the \snia\ age distribution, we adopt a smooth functional form for our ``nominal'' \snia\ DTD as follows:
\begin{equation}
  \phi(t) \propto \frac{\left(t/t_p\right)^{\alpha}}{\left(t/t_p\right)^{\alpha-s}+1}
\label{eq:smooth_dtd}
\end{equation}
At late times ($t \geq t_p$), this function form rapidly approaches a $t^{s}$ power law, with observations favoring a power law slope for the \snia\ DTD of $s=-1$.  At early times ($t \leq t_p$) this function is a high order ($\alpha$) polynomial which flattens to zero rapidly below some characteristic ``prompt'' timescale ($t_p$).  This functional form for the \snia\ DTD is of course entirely artificial, but reflects the qualitative traits inferred from observations.  We will show in Section~\ref{sec:dtd_uncertainties} that our primary results are insensitive to the exact form of the DTD, but instead are driven very strongly by the steep ($\sim t^{-1}$) slope of the DTD.  We adopt a fiducial DTD with $s=-1$ and $t_p=0.3$~Gyr for the main analyses of this paper.

% FIGURE
\begin{figure*}
\begin{center}
\includegraphics[width=0.90\textwidth]{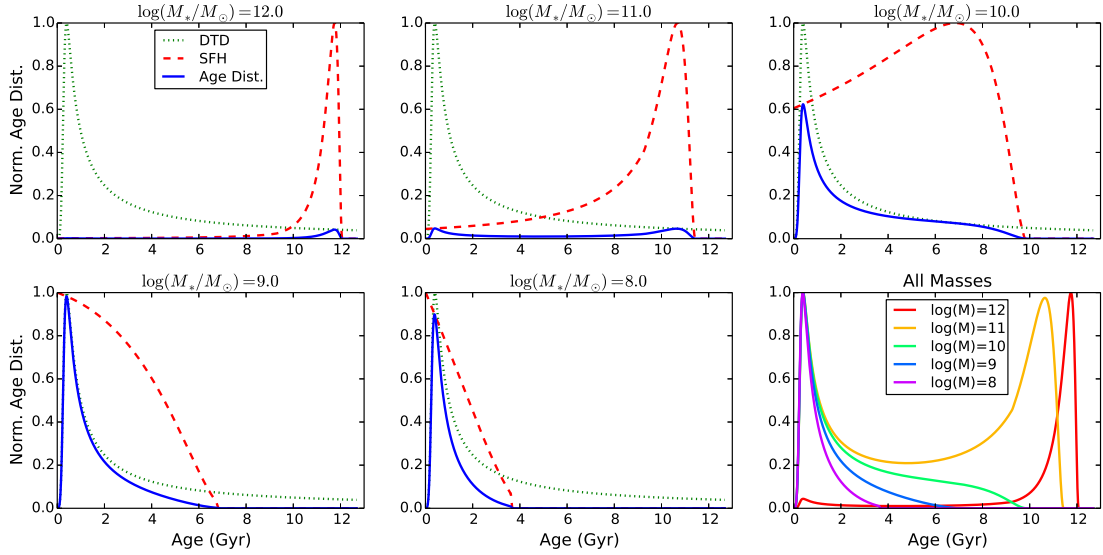}
\caption{\snia\ age distributions (solid blue curve in the first five panels) calculated as the integrand of the convolution of the \snia\ DTD (dotted green curve) and galaxy SFH from our empirical models (dashed red curve).  The SFHs have been plotted over lookback time to reflect the convolution step, and the age distribution is the product of the DTD and SFH.  The bottom right panel shows the peak-normalized age distribution for the chosen galaxy mass scales presented in the first five panels.}
\label{fig:age_dists}
\end{center}
\end{figure*}

% DTD + SFH convolution results!
In Figure~\ref{fig:age_dists} we show the age distribution calculated by convolving our smooth DTD function (Equation~\ref{eq:smooth_dtd}) with the mean SFHs from Section~\ref{sec:mass_buildup} for several values of stellar mass. The bottom right summary panel of this figure compares the \snia\ age distributions for the various host masses, and illustrates one of the key results of this paper: {\em in galaxies that are actively star-forming, the \snia\ age distribution is shaped more strongly by the shape of the DTD than by that of its host galaxy SFH}. For the three example galaxies in Figure~\ref{fig:age_dists} whose SFH has strong recent star formation, the \snia\ age distributions all peak at the same age where the DTD peaks. This is because the evolution of SFR in these galaxies is much more gradual than the steep decline in \snia\ rates as a function of stellar age.

% passive galaxies!
The two example galaxies in Figure~\ref{fig:age_dists} with SFH dominated by an old stellar population exhibit a markedly different \snia\ age distribution. In these examples, the galaxies rapidly form a large mass of stars over a short duration ($\sim$1~Gyr) at a distant past epoch ($\sim$8-10~Gyr). Here the resultant \snia\ age distribution mimics the stellar age distribution (note that $t/\Delta t$ is large so the $t^{-1}$ DTD does not significantly alter the shape of the host stellar age distribution). We note that for galaxies with even a small amount of recent star formation (\logm$=11$ in Figure~\ref{fig:age_dists}), a non-negligible component of the \snia\ age distribution appears at the DTD peak. However, the second major conclusion of this work holds true: {\em in galaxies dominated by old stellar populations, the mean \snia\ age traces the mean epoch when the stellar mass was formed}.

%------------------------------------------------------
\subsection{The Bimodality of \snia\ Ages}
\label{subsec:age_bimodality}
\begin{figure*}
\begin{center}
\includegraphics[width=0.90\textwidth]{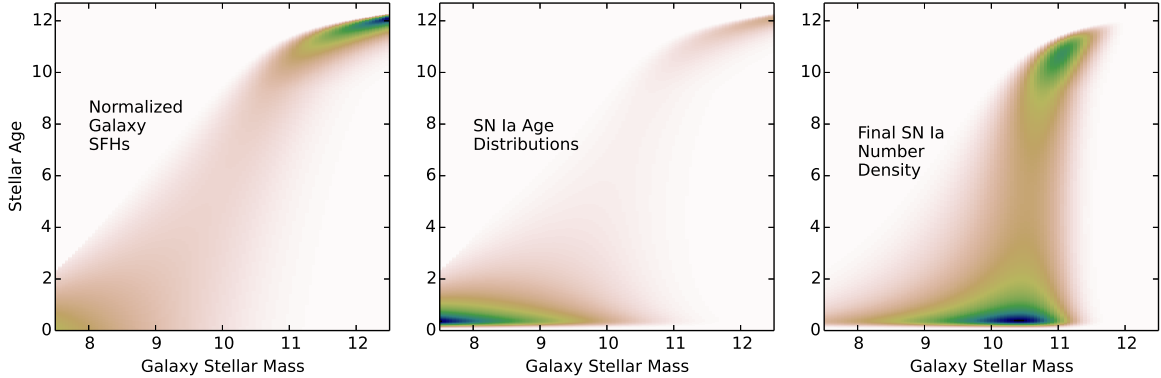}
\caption{{\em Left:} Normalized mean galaxy SFHs as a function of total stellar mass. {\em Middle:} \snia\ age distribution versus host galaxy mass (not re-normalized: age distribution per unit mass). {\em Right:} Intrinsic distribution of \sneia\ in progenitor age-host mass space in the local Universe ($z=0$).}
\label{fig:bimodality_origin}
\end{center}
\end{figure*}

%------------------------------------------------------
With our galaxy mass assembly models in hand and a chosen \snia\ DTD, we can calculate the intrinsic distribution of progenitor ages for \sneia\ occurring in the local (and distant) Universe. We previously calculated the average (normalized) SFH as a function of galaxy stellar mass, which we plot in two-dimensional age-mass parameter space in the left panel of Figure~\ref{fig:bimodality_origin}.  Convolved with our nominal \snia\ DTD (Eq.~\ref{eq:smooth_dtd}), this produces the relative age distributions, which we show in age-mass space in the middle panel of Figure~\ref{fig:bimodality_origin} (note these are {\em not} re-normalized). 

% FIGURE

% Schechter function
Finally, to obtain the intrinsic distribution of \sneia\ in age-mass space, we need to account for the volume density of total stellar mass in each logarithmic galaxy mass bin. Again, this is simply a modified Schechter function, for which we use the low redshift values from \citet{moustakas13} as previously noted. The final number density of \sneia\ in progenitor age-host mass space is shown in the right panel of Figure~\ref{fig:bimodality_origin}.

% bimodality discussion
The age distribution of \sneia\ in Figure~\ref{fig:bimodality_origin} shows a striking existence of two primary modes. The young component of this distribution, commonly referred to as the ``prompt'' component in the literature, is dominant at low stellar mass and exhibits the same peak age in all actively star-forming galaxies. The old component, often called the ``tardy'' component, forms a subdominant component at old stellar ages almost exclusively in high-mass galaxies.

% A+B discussion
Figure~\ref{fig:bimodality_origin} also naturally explains why the \snia\ rate has been so well parametrized by the two-component (or ``A+B'') model, with one component of the rate proportional to galaxy stellar mass (``A'') and another proportional to its SFR (``B''). Massive galaxies formed their stars during a short period in the distant past, so the relative \snia\ rate (corresponding to the DTD at that age) depends on the total stellar mass of the galaxy since all the stellar mass was formed at a similar epoch. Actively star-forming galaxies (regardless of their detailed SFH) are dominated by the DTD peak age, which closely tracks the recent galaxy SFR.

\begin{figure}
\begin{center}
\includegraphics[width=0.45\textwidth]{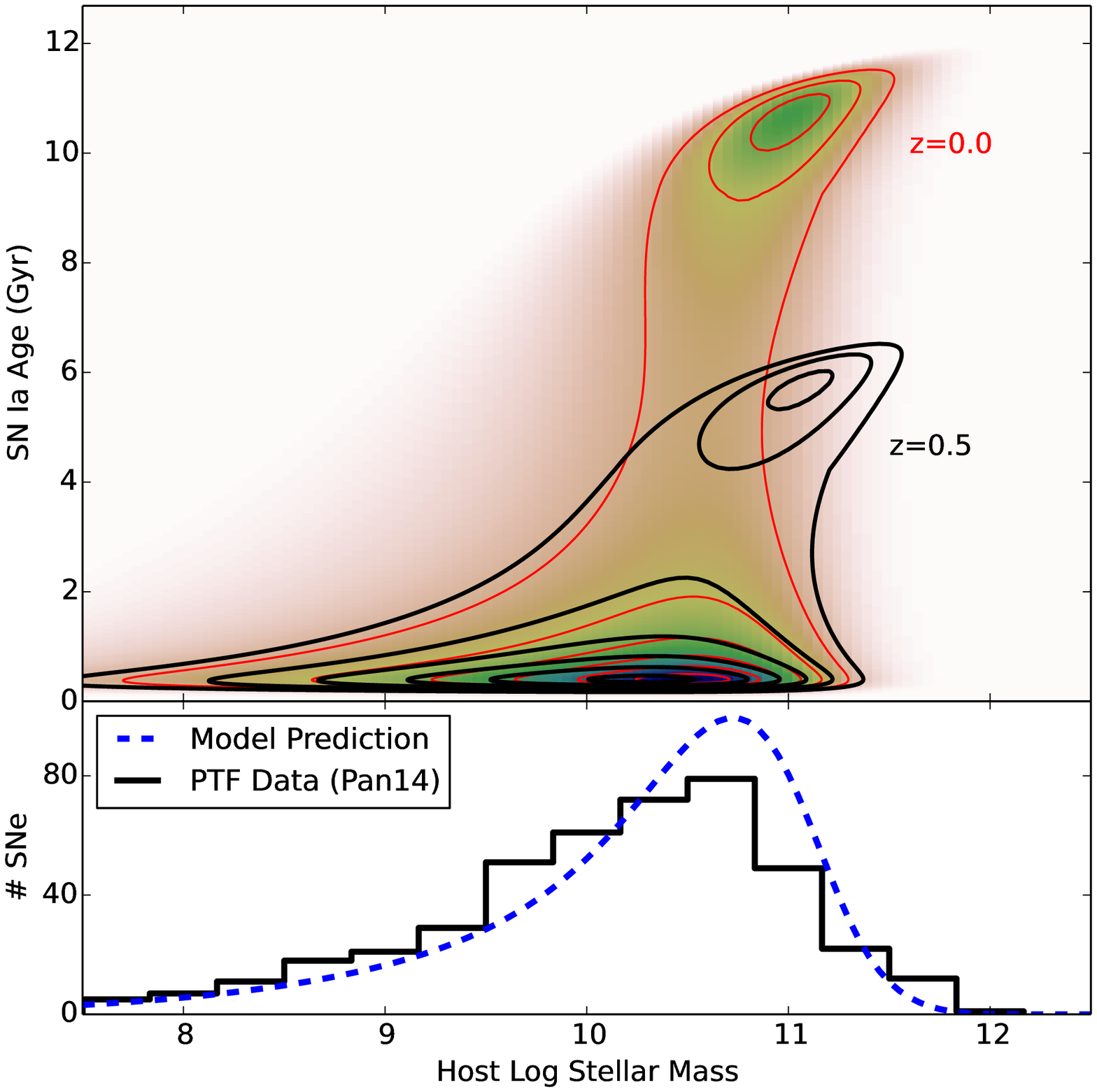}
\caption{Top: Same as right panel of Figure~\ref{fig:bimodality_origin}, but with contours drawn for both $z=0.0$ (thin red) and $z=0.5$ (thick black). Bottom: \snia\ host galaxy mass distribution as predicted by our models at $z=0$ (blue curve) compared to data (black histogram) from PTF \citep{pan14}.}
\label{fig:bimodality_evolution}
\end{center}
\end{figure}

%---------------------------
% bimodality evolution
Because our galaxy mass assembly models are tracked self-consistently from high redshift, we can inspect the SFHs of galaxies of any stellar mass at any redshift. This allows us to inspect how the age distribution of \snia\ ages evolves over cosmic time. As an example, we recalculate the mean SFHs for galaxies at redshift $z=0.5$, and use the Schechter function parameters at that redshift \citep[the sum of the blue and red galaxy mass functions fits from][]{borch06}, to derive the \snia\ distribution in age-mass space at $z=0.5$. We show this $z=0.5$ age distribution as the thick contours in Figure~\ref{fig:bimodality_evolution}, compared to the $z=0.0$ distribution (density plot and thin red contours).

% FIGURE

Several key features are evident from this result. First, the bimodality of \snia\ ages persists even at high redshift, due to the nature of galaxy SFHs generally corresponding to either current active star formation or quenched past star formation. Next, the SFHs of star-forming galaxies at high redshift are again shallow compared to the $t^{-1}$ \snia\ DTD, meaning the prompt \snia\ component again arises predominantly from the age of the DTD peak. This held true at every redshift we calculated, implying the critical result {\em young (prompt) \sneia\ arise from the same uniform progenitor age group throughout cosmic history}.

Finally, critical contrast is seen for the old (tardy) \snia\ component. Though the old component again arises from the past epoch when star formation occurred rapidly and then ceased in massive galaxies, this occurred at a different past age for $z=0.5$ galaxies than for $z=0.0$ galaxies. This means the {\em old \sneia\ arise from different progenitor ages at different redshifts}. Thus, \sneia\ in passive galaxies present a diverse and evolving progenitor age group over cosmic history, while young \sneia\ from actively star-forming galaxies remain highly uniform in their progenitor ages at all cosmic epochs.

%---------------------------
% host mass distribution
As an additional cross-check against data, we show in the lower panel of Figure~\ref{fig:bimodality_evolution} the prediction of our fiducial models for the \snia\ host galaxy mass distribution compared to that observed at low redshift by PTF \citep{pan14}.  The shapes of the predicted and observed distributions are modestly consistent, but with a mild over-prediction of high-mass hosts.  We will show in future work (Childress \& SNfactory, in prep.) that the shape of this host mass distribution can actually {\it constrain} the \snia\ DTD.

%------------------------------------------------------
\subsection{Evolution of the \snia\ Age Distribution}
\label{subsec:age_dist_vs_z}
%------------------------------------------------------
Following the same procedure for calculating the \snia\ age distribution in a single galaxy, we can couple our DTD to the cosmic SFH (again from B13) to calculate the global \snia\ age distribution at a given redshift. This age distribution can also be calculated by integrating the \snia\ age distribution as a function of galaxy mass, weighted by the modified Schechter-function (i.e., the right panel of Figure~\ref{fig:bimodality_origin}). However, as we showed in Figure~\ref{fig:models_vs_csfh}, our baseline models do not exactly match the observed CSFH (though see Section~\ref{subsec:fine_tuned_model}), so we prefer the B13 CSFH for calculating the aggregate \snia\ age distribution.

Note this the \snia\ age distribution is a different quantity than the delay time distribution, which reflects the intrinsic progenitor age distribution for all \sneia\ produced by a single burst of star formation, counted over all epochs following the SF burst. The age distribution we discuss here is the distribution of progenitor ages which give rise to the sample of \sneia\ occurring at a single epoch of cosmic time, summed over all galaxies in a large homogeneous volume of space (note this age distribution {\em can} take the same form as the DTD if produced from galaxies having a constant SFR for infinite duration).

% FIGURE
\begin{figure}
\begin{center}
\includegraphics[width=0.45\textwidth]{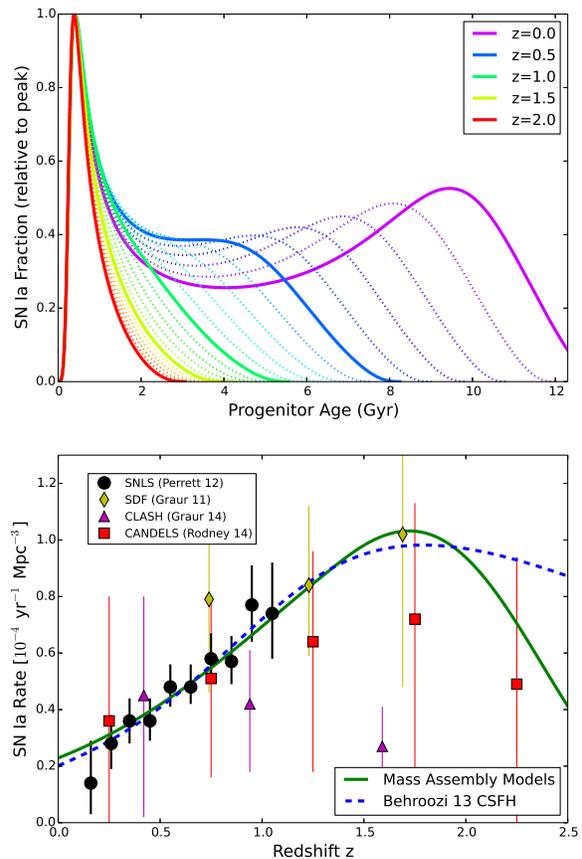}
\caption{Top: Intrinsic \snia\ age distribution as a function of redshift from $z=0.0$ (purple) to $z=2.0$ (red) in steps of 0.1 (dotted lines), with half-integer redshift steps denoted as thick solid lines. Based on convolution of the B13 CSFH with our smooth DTD (Equation~\ref{eq:smooth_dtd}). Bottom: Predicted evolution of the \snia\ rate with redshift for our mass assembly models (solid green curve) and the B13 CSFH (dashed blue curve) compared to data from SNLS \citep{perrett12}, Subaru Deep Field \citep[SDF;][]{graur11}, CLASH \citep{graur14}, CANDELS \citep{rodney14}.}
\label{fig:age_dist_evolution}
\end{center}
\end{figure}

% present age dist
In Figure~\ref{fig:age_dist_evolution} we present this age distribution for \sneia\ at various redshifts from $z=0.0$ to $z=2.0$.
%old component gets younger at higher z
The age distributions in Figure~\ref{fig:age_dist_evolution} exhibit several important features. First, the bimodality of \snia\ ages is evident up to redshift $z=0.5$, but the mean age of the old (tardy) component decreases at higher redshift and eventually the two components become blended. Second, the prominence of the tardy peak decreases with redshift, but not so extremely as expected from ``A+B'' rate calculations coupled to the evolution of the cosmic star formation rate. Using the $z=0$ ``A'' and ``B'' coefficients would imply that the fraction of old (tardy) \sneia\ drops by a significant factor by redshift $z=0.5$ \citep{sullivan06, howell07}, but the younger age of the high-redshift tardy component results in a higher rate than the ``A+B'' calculation due to the $t^{-1}$ DTD. This critical result implies that {\em the old (tardy) component of \sneia\ comprises a non-negligible fraction of \sneia\ at the redshifts ($z\sim0.5$) of existing major cosmological \snia\ samples.}

%------------------
% rate vs. z!!
As a further consistency check, we can predict the evolution of the integrated \snia\ rate with redshift using our galaxy mass assembly models.  In the lower panel of Figure~\ref{fig:age_dist_evolution}, we plot the prediction of our models for the \snia\ rate as a function of redshift, as well as the prediction for the observed CSFH from B13.  These predicted rates are compared to observations from several key recent \snia\ rates studies from major surveys: the Supernova Legacy Survey \citep[SNLS;][]{perrett12}, the Subaru Deep Field \citep[SDF;][]{graur11}, and CLASH survey \citep{graur13}, and the CANDELS survey \citep{rodney14}.  Both models are normalized to the SNLS \citep{perrett12} rate data, and show good agreement with the redshift evolution of the \snia\ rates from this precise analysis.

%%%%%%%%%%%%%%%%%%%%%%%%%%%%%%%%%%%%%%%%%%%%%%%%%%%%%%%%%%%%%%%%%%%%%%%%%%%%%
\section{\snia\ Ages Along the Galaxy Mass Sequence}
\label{sec:age_vs_mass}
\begin{figure}
\begin{center}
\includegraphics[width=0.45\textwidth]{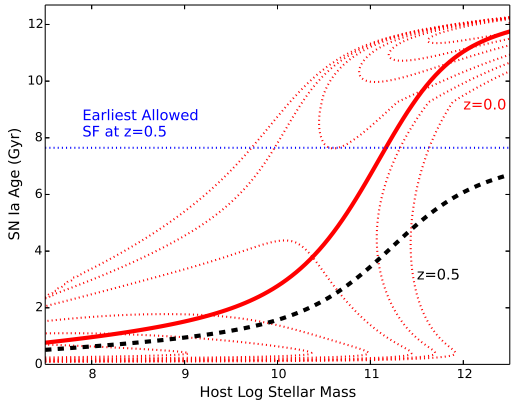}
\caption{Mean age of \sneia\ as a function of host galaxy mass at redshift $z=0$ (solid red curve) and $z=0.5$ (dashed black curve). For reference, we show the \snia\ age distribution at $z=0$ as the dotted red contours (normalized at each galaxy mass value), as well as the earliest epoch of allow star formation at $z=0.5$ (dotted blue line).}
\label{fig:ages_sf_times}
\end{center}
\end{figure}

%---------------------------
In addition to calculating the aggregate \snia\ progenitor age distribution across all galaxies, we can examine how the \snia\ age distributions of Section~\ref{sec:age_dists} vary as a function of host galaxy stellar mass. In Figure~\ref{fig:ages_sf_times}, we plot the mean of the \snia\ age distribution as a function of host galaxy stellar mass both in the local ($z=0.0$) and distant ($z=0.5$) Universe.

% FIGURE

% comment on results 
% - t^-1 makes age follow SF start!
From Figure~\ref{fig:ages_sf_times} we see a clear trend of \snia\ age with host galaxy mass that broadly reflects the behavior identified from \snia\ age distributions of individual galaxies in Section~\ref{subsec:snia_age_dist}.  Low-mass galaxies are actively star-forming, and thus have \snia\ age distributions strongly dominated by younger progenitor populations.  In contrast, high-mass galaxies formed most of their stars in the distant past, so the \sneia\ they produce are consistently old.  The age-mass trend shows a rapid transition between $10.0\leq$\logm$\leq11.5$ as galaxies are influenced and then dominated by quenching.
% - looks like HR step!
This age transition resembles the step-like structure of the \snia\ Hubble residual trend with host galaxy mass, as identified by C13 and \citet{johansson13}.

%------------------
% X1 VS. MASS

% FIGURE
\begin{figure}
\begin{center}
\includegraphics[width=0.45\textwidth]{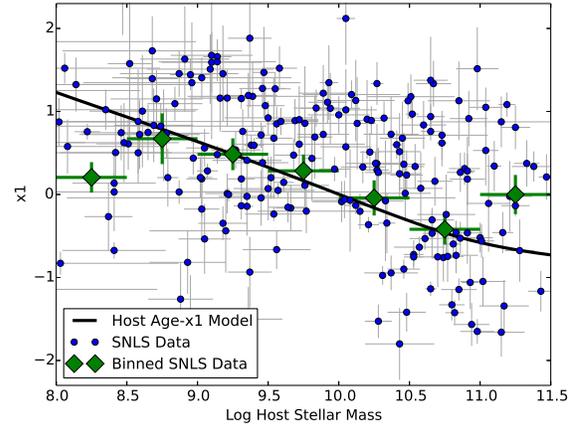}
\caption{Observed \snia\ stretch (here SALT2 $x1$) versus host galaxy mass for the SNLS \citep{howell09, sullivan10} sample, both as individual objects (blue points) and binned averages (green diamonds) in bins of 0.5~dex in host stellar mass.  To this we compare the prediction for mean $x1$ versus host mass using the relationship between \snia\ stretch and host galaxy age \citep{johansson13} coupled to our galaxy mass assembly models.}
\label{fig:x1_mass_trend}
\end{center}
\end{figure}

% x1-mass relation discussion
As an additional consistency check, we take advantage of the fact that our models predict the mean age of galaxies as a function of their stellar mass.  Host galaxy age is also known to tightly correlate with \snia\ stretch \citep{howell09, neill09, johansson13}, allowing us to predict the mean stretch of \sneia\ versus host galaxy mass.  In Figure~\ref{fig:x1_mass_trend} we show the stretch (here SALT2 $x1$) versus mass data from SNLS \citep{howell09, sullivan10}, both as individual SNe and in bins of 0.5~dex in host mass.  To this we compare the prediction for $x1$ versus host mass using the $x1$-age (note here this is mean host galaxy age, not the age of the SN) slope from \citet{johansson13} with a zeropoint that best fits the data.  This model shows surprisingly good agreement with the observed evolution of $x1$ with host mass, lending support to our galaxy mass assembly models.

%------------------
% EVOLUTION WITH REDSHIFT!
Our results for mean \snia\ age versus host mass support the suggestion by C13 that the step-like shape of the \snia\ Hubble residual trend with mass is driven by progenitor age. The evolution of galaxy populations over cosmic time implies this age-mass trend will also evolve with redshift. In Figure~\ref{fig:ages_sf_times}, we also show the \snia\ mean age trend with host galaxy mass at $z=0.5$, clearly illustrating this effect

% FIGURE
\begin{figure}
\begin{center}
\includegraphics[width=0.45\textwidth]{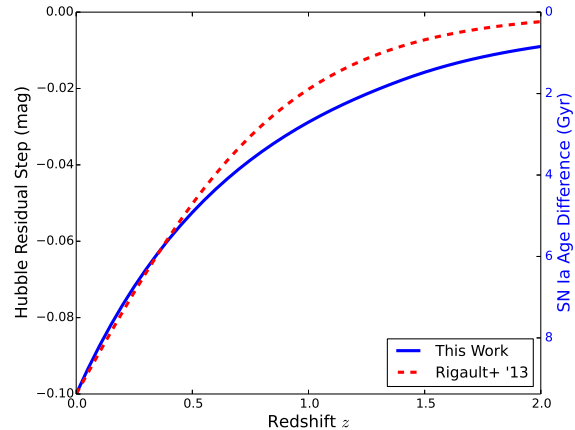}
\caption{Evolution of the \snia\ Hubble residual (HR) ``mass step'' value with redshift, from our SFH models (solid blue curve) and using Equations (5) and (6) from \citet{rigault13}.  For illustrative purposes, we set the HR step to -0.10~mag at $z=0$, and scale this value by the mean age difference between high and low-mass bins (split at \logm$=10$).}
\label{fig:hr_trend_vs_z}
\end{center}
\end{figure}

To quantify the evolution of the \snia\ age-mass trend and the resulting Hubble residual step in host mass (hereafter, the ``HR step''), we calculate the mean SFH for galaxies at the same mass sampling as above, for all redshifts between $z=0$ and $z=2$ in steps of $\Delta z=0.05$.  At each redshift, we calculate the difference between mean age in the high-mass bin (defined by \logm$\geq10$) and low-mass bin.  We find the age difference between mean \snia\ age at \logm$=12$ and \logm$=8$ scale similarly to this binned age difference.  We also find that the galaxy mass scale at which mean \snia\ age was halfway between the high- and low-mass values (i.e. the ``transition'' mass) was consistent to within 0.1~dex across all redshifts.

We examine the evolution of the HR step with redshift under the scenario where it is proportional to age differences.  In Figure~\ref{fig:hr_trend_vs_z} we plot the HR step versus redshift assuming the $z=0$ value is -0.10~mag.  We see that the HR step decreases significantly at high redshifts.  Similarly, R13 find that the HR step appeared to be driven by \sneia\ in passive environments in high-mass hosts, and calculated the HR step evolution with redshift assuming it scales with the cosmic sSFR.  We show the results of the R13 model in Figure~\ref{fig:hr_trend_vs_z} compared to ours, and they appear quite similar.  One distinct difference is that the R13 model predicts the HR step is nearly zero at $z=2$, while our models still exhibit a nonzero value.  This is due to our models tracking the diminishing, but non-vanishing, age difference between the prompt \snia\ timescale (set consistently by the DTD peak) and the tardy \snia\ timescale (set by the evolving stellar ages in massive galaxies).

%%%%%%%%%%%%%%%%%%%%%%%%%%%%%%%%%%%%%%%%%%%%%%%%%%%%%%%%%%%%%%%%%%%%%%%%%%%%%
\section{\snia\ Ages: Sensitivity to Galaxy SFHs}
\label{sec:sfh_uncertainties}
%---------------------------
The buildup of stellar mass in galaxies is a rich field of study, and a full census of possible approaches to modeling galaxy mass assembly is beyond the scope of this work. Instead, we are concerned with how variations in the nature of our mass assembly models affect the ages of \sneia. Our models were used to describe how the age distribution of stars in the local Universe is distributed along the galaxy mass sequence, which allows us to calculate the trend of \snia\ ages with host galaxy mass (Section~\ref{sec:age_vs_mass}). In this Section we will thus examine how subtly different parametrizations of galaxy mass assembly affect the \snia\ progenitor age-host mass trend.

\begin{figure}
\begin{center}
\includegraphics[width=0.45\textwidth]{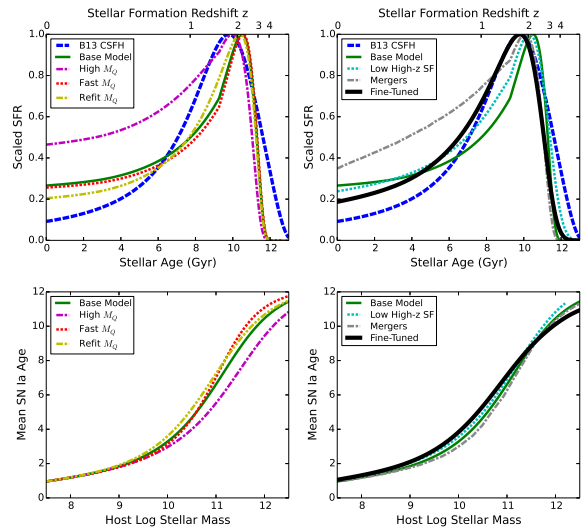}
\caption{Top panels: Cosmic SFH predicted by each variation of our mass assembly models compared to the observed cosmic SFH from \citet{behroozi13}. Bottom panels: \snia\ age versus host mass trend at $z=0$ for each variation of the mass assembly models. Note that even mass assembly modifications that produce seemingly major changes to the CSFH ultimately produce minor changes to the \snia\ age versus host mass trend.}
\label{fig:sfh_var}
\end{center}
\end{figure}

% summary of subsections and introduction of figure
As an additional diagnostic to test the reliability of the varied mass assembly models presented in this Section, we test the predicted cosmic SFH (CSFH) for each model against the observed CSFH (as in Figure~\ref{fig:models_vs_csfh}).  In the top panels of Figure~\ref{fig:sfh_var} we show the predicted CSFH for each model variation compared to the observed CSFH of B13. The bottom panels of Figure~\ref{fig:sfh_var} show the \snia\ age versus host mass trend predicted by the model variations. Section~\ref{subsec:quenching_variations} presents variations to the galaxy SF quenching prescription, Section~\ref{subsec:smz_variation} discusses the effect of an SMz relation which gradually plateaus at high redshift, while Section~\ref{subsec:mergers} examines the possible impact of galaxy mergers. A ``fine-tuned'' model incorporating multiple adjustments is presented in Section~\ref{subsec:fine_tuned_model}, while a summary of all the model variations is presented in Section~\ref{subsec:sfh_var_summary}. The equations for the modified parametrizations employed in this Section are presented in Appendix~\ref{app:equations}.

% FIGURE

%------------------------------------------------------
\subsection{Variation in SF Quenching}
\label{subsec:quenching_variations}
%------------------------------------------------------
Quenching of galaxy star formation is perhaps the most uncertain property of galaxy evolution incorporated into our models. The observational distinction between active and passive galaxies is typically performed with broad-band photometric colors, but by necessity this technique typically probes different rest wavelengths at different redshifts. As a result, several possible variations to the true quenching of galaxy SF could be possible, and we examine several of those here.

% higher queching mass
{\it Underestimated Quenching Masses?} First we consider whether the quenching mass scale estimated from photometric colors may not correspond directly to the epoch when SF ceases. Specifically, we investigate whether the true quenching mass may be underestimated by some amount. In the left panels of Figure~\ref{fig:sfh_var} we present the results of repeating our galaxy mass assembly procedure, but with a quenching mass scale that is 0.5~dex higher than the \citet{muzzin13} parametrization employed in our nominal models. The result shows that the recent SF in the Universe is increased, since $L^*$ galaxies (which contribute the most to SF at $z<1$ in our models) are not penalized for quenching. For \snia\ ages, this effect slightly raises the transition mass scale for \snia\ ages, but the clear age transition with host mass is robustly maintained.

% narrow quenching
{\it Faster Transition To Quenched?} Next we examine the effect of shortening the galaxy mass scale over which galaxies transition from active to passive. In our nominal models this was 1.5~dex, and in Figure~\ref{fig:sfh_var} we present the results of reducing this to 1.1~dex \citep[closer to the value implied from the GAMA mass functions of][]{baldry12}. This has an almost negligible affect on the predicted CSFH, and merely causes the \snia\ age-mass transition to occur on a proportionally shorter mass scale.

% alternate quenching function
{\it Different $M_Q(z)$ Parametrization?} Finally, we test the effect of employing a different parametrization for the quenching mass as a function of redshift $M_Q(z)$. We use the \citet{muzzin13} data points up to redshift $z=2$, along with the $z=0.06$ value for GAMA from \citet{baldry12}, as well as the $z\leq2$ values from the ZFOURGE survey mass functions from \citet{tomczak14}. We find these data were consistent with a linear functional form for $M_Q(z)$, resulting in a generally lower quenching mass below $z\leq1$ than that predicted in the \citet{muzzin13} parametrization. The CSFH below $z\leq1$ in turn shows better agreement with the observed CSFH from B13, and the \snia\ age-mass trend is virtually unchanged.

% nonzero sfr when quenched
%{\it SF Never Fully Quenched?} stuff

%------------------------------------------------------
\subsection{Impact of High-z SF Intensity}
\label{subsec:smz_variation}
%------------------------------------------------------
The SMz relation we employ is based on the Z12 parametrization which describes observed data very well up to $z\sim2$. However, our extrapolation of this relation beyond $z=2$ results in very intense high redshift SF (since the Z12 parametrization increases monotonically with redshift). Consequently, massive galaxies in our models form the majority of their stars in an extremely rapid period of time at high redshift, as they rapidly reach the quenching mass. This results in a narrow time scale for the formation of most of the Universe's stellar mass in our models, seen as the narrowness of our model CSFH compared to the B13 data.

To examine a possible remedy to this situation, we construct an alternate model for the redshift evolution of the SMz relation while retaining the Z12 mass dependence. Our alternate parametrization (whose formulae are presented in Appendix~\ref{app:equations}) is tuned to mimic the Z12 redshift dependence below $z=2$, but to have a gentler redshift dependence above this redshift. This slowing of the high-redshift evolution of the SMz relation is observationally motivated by a slower evolution of galaxy SFRs above redshift $z=2$ \citep{stark13}. The resulting models present a CSFH which is broader in age (top right panel of Figure~\ref{fig:sfh_var}), and thus closer to the observed CSFH. The \snia\ age-mass trend retains its previously noted qualitative characteristics.

%------------------------------------------------------
\subsection{The Effect of Galaxy Mergers}
\label{subsec:mergers}
%------------------------------------------------------
% merger intro
The galaxy mass assembly models we have presented thus far assume that galaxies undergo secular evolution, where a single galaxy evolves independently of every other galaxy in the Universe. This neglects the observational fact that galaxies often undergo mergers, and this could contribute significantly to the buildup of stellar mass in galaxies especially since $z=1$ as the CSFH continuously declines \citep{bell06, conselice08, deravel09}, particularly along the red sequence \citep{faber07}. Indeed, recent work from the GAMA survey \citep{robotham14} has shown that while star-formation dominates stellar mass assembly at low galaxy mass scales (less than \logm$\sim10.6$), mergers represent the dominant mode through which massive galaxies continue to build their stellar mass.

The exact contribution of mergers to galaxy mass assembly is observationally difficult to constrain, and has only been examined across all galaxy mass scales in the local Universe \citep{robotham14}. Estimating merger rates is further complicated by the necessity to combine the volumetric merger fraction measured from a snapshot of some epoch of the Universe with some unknown timescale for the duration of the mergers. Indeed, examination of mergers in the Millennium Simulation \citep{millennium} has suggested that mergers do not contribute the majority of stellar mass growth in galaxies, and therefore secular evolution must be the dominant effect \citep{genel08}. Newer simulations which also account for detailed baryon physics \citep{vogelsberger14, genel14} will likely shed further light on the importance of mergers for galaxy mass assembly at all redshifts.

For the purposes of our \snia\ age analysis, we need to investigate what effect mergers could have on the distribution of stellar ages across the galaxy mass sequence. To test this, we repeat our galaxy mass assembly process for select tracks and introduce mergers to the process. Specifically, at each time step we calculate a random likelihood of a merger occurring with a total rate of 0.3~Gyr$^{-1}$. We then add the SFH of the building ``parent'' galaxy to its merging ``child'' galaxy, assuming the child has undergone secular evolution up to the point of merger. The mass of the child galaxy being merged is chosen from an exponential distribution with mean value 0.2 times the mass of the parent galaxy. After the merger the merged galaxy continues to undergo secular evolution following the SMz relation and quenching prescription as outlined in Section~\ref{sec:mass_buildup} (with additional mergers allowed).

% FIGURE
\begin{figure}
\begin{center}
\includegraphics[width=0.45\textwidth]{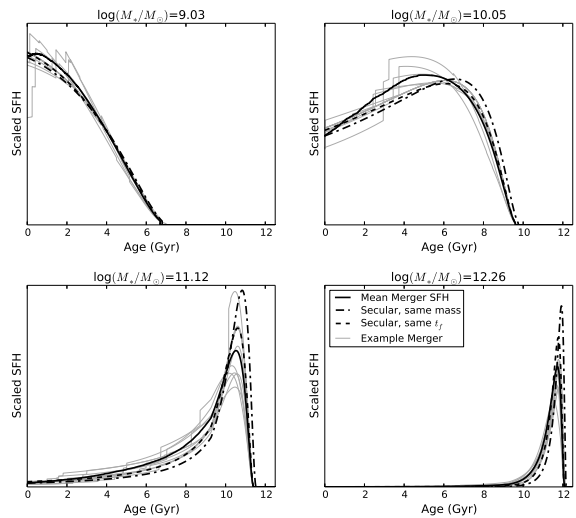}
\caption{Mass assembly models incorporating mergers for four fixed formation times corresponding to \logm\ $=9,10,11,12 M_\odot$ in the secular models.  The mean merger track SFHs are shown as the thick solid black line, while single merger SFHs are thin grey lines.  The dashed line represents the secular SFH for the same formation time $t_f$, while the dashed-dotted line represents the secular SFH with the same final mass as the mean merger SFH mass.}
\label{fig:merger_results}
\end{center}
\end{figure}

In Figure~\ref{fig:merger_results} we show the merged SFHs for four of the selected mass assembly tracks, whose secular tracks correspond to integer intervals in log stellar mass.  For each track, we repeat 100 random merger simulations (a subset of which are shown as thin grey lines in each panel), and calculate the mean SFH and mean final mass of the merged galaxies. 

Repeating this process for numerous tracks, we again arrive at a prescription for the mean SFH of galaxies as a function of their total stellar mass. With this we can calculate the predicted CSFH and \snia\ age versus host mass trend. Our merger tracks show an excess of stars at young ages compared to the CSFH predicted by our fiducial models, which further increases the discrepancy with the observed CSFH. This is because the merger effect causes galaxies to have younger average ages for a given stellar mass when compared to the SFH from secular evolution. This also results in slightly younger \snia\ ages at transitional host galaxy mass scales ($10^{10}-10^{11}M_\odot$), but the \snia\ age transition is qualitatively quite consistent with our base model.

% does merger effect on SFR already appear in the measured SMz?

%------------------------------------------------------
\subsection{A Fine-Tuned Mass Assembly Model}
\label{subsec:fine_tuned_model}
%------------------------------------------------------
% introduce fine-tuned model
As some variations to our mass assembly models showed a more favorable agreement with the observed CSFH, we construct a final set of models which incorporates several of these effects. This ``fine-tuned'' model employs the refit quenching versus redshift relation, the decreased (plateau) high-redshift SMz parametrization, and an additional alteration to the quenching prescription which we now describe.

High-redshift galaxy stellar mass functions have revealed that actively star-forming galaxies may be a non-vanishing component of galaxy populations at {\em all} galaxy mass scales. Furthermore, many galaxies which evolve off the star-forming main sequence exhibit some residual SF activity \citep{schawinski09}, especially those galaxies which quench SF by slowly exhausting their gas reservoir \citep{schawinski14}.

To capture this behavior, we enforce a redshift-dependent minimum active galaxy fraction, which manifests as a minimum allowable value for the quenching penalty function (see Appendix~\ref{app:equations} for details). As an example, this minimum SF galaxy fraction at $z=2$ is 35\%, meaning even the most massive galaxies will retain a SFR of 35\% their SMz value. This results in more massive galaxies building their stellar mass over a more extended period of time, resulting in a favorable agreement of the CSFH prediction compared to observations (see Figure~\ref{fig:sfh_var}). Because this ``minimum SF fraction'' effect is manually constructed to account for a poorly constrained observational effect, we denote this set of models as the ``fine-tuned'' models.

%------------------------------------------------------
\subsection{SFH Variations: Summary of Results}
\label{subsec:sfh_var_summary}
%------------------------------------------------------
% summarize all the models
We explored various alterations to the nominal galaxy mass assembly prescriptions, which show improved agreement with the observed CSFH in some cases and worse agreement in others. Regardless of the galaxy mass assembly details, the transition of \snia\ ages from young SNe in low-mass galaxies to old SNe in high-mass galaxies is preserved in all variations of our models.
% conclusion
This is because it is impossible to avert the partitioning of old stars to massive galaxies and young stars to low-mass galaxies. This well-known ``downsizing'' \citep{cowie96, faber07} of galaxy stellar mass and star-formation is driven by the relationships between sSFR and quiescent galaxy fraction with stellar mass. Consequently, \snia\ ages are affected by the mass assembly history of their host galaxies, and the \snia\ age transition with host mass is unavoidable.

%%%%%%%%%%%%%%%%%%%%%%%%%%%%%%%%%%%%%%%%%%%%%%%%%%%%%%%%%%%%%%%%%%%%%%%%%%%%%
\section{\snia\ Ages: Sensitivity to the \snia\ DTD}
\label{sec:dtd_uncertainties}
%---------------------------
In this Section we inspect how the \snia\ age distribution changes for different forms of the \snia\ DTD. For simplicity and visual clarity, we will focus on how changes in the \snia\ DTD affect the aggregate progenitor age distribution of \sneia\ in the local ($z=0.0$) Universe.

% FIGURE
\begin{figure}
\begin{center}
\includegraphics[width=0.45\textwidth]{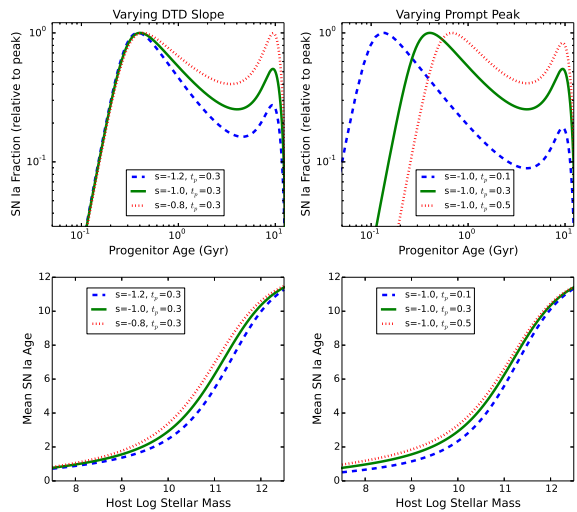}
\caption{Changes in the \snia\ progenitor age distribution at $z=0.0$ (top panels) and \snia\ age versus host mass trend (bottom panels) resulting from changes in the \snia\ DTD. Left panels: varying the power law slope $s$ from Equation~\ref{eq:smooth_dtd}. Right panels varying the prompt timescale $t_p$ of the \snia\ DTD from Equation~\ref{eq:smooth_dtd}.}
\label{fig:func_dtd_var}
\end{center}
\end{figure}

We begin with altering the general shape of the \snia\ DTD by varying the parameters in Eq.~\ref{eq:smooth_dtd}. In Figure~\ref{fig:func_dtd_var}, we show the results of varying the power law slope $s$ of the DTD (top panel) and the prompt timescale $t_p$ (bottom panel). Varying the power law slope of the DTD has no effect on the peak age of the prompt component, but has a significant effect on the relative ratio of the prompt and tardy components of the \snia\ progenitor age distribution. Varying the prompt timescale changes not only the peak age of the prompt component (by construction), but also the ratio of prompt to tardy \sneia\ due to the increased relative rate when the prompt component is relatively younger.

\begin{figure}
\begin{center}
\includegraphics[width=0.45\textwidth]{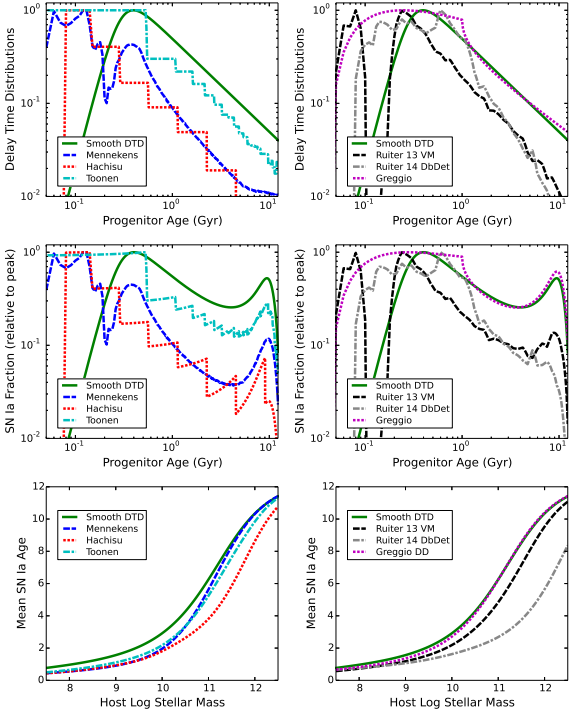}
\caption{Changes in the \snia\ progenitor age distribution at $z=0.0$ (middle panels) and \snia\ age versus host mass trend (bottom panels) resulting from the use of various literature DTDs (top panels).}
\label{fig:lit_dtd_var}
\end{center}
\end{figure}

% LITERATURE DTDs
While the simple DTD used throughout this work is intentionally artificial, many binary population synthesis studies have been conducted to produce physically realistic DTDs for given progenitor evolution scenarios. We calculate the \snia\ age distributions at $z=0$ for select DTDs from the literature coupled to the B13 CSFH. In Figure~\ref{fig:lit_dtd_var} we show the result for the DD scenario of \citet{mennekens10} with $\alpha=1.0$ and $\beta=1.0$, the combined SD DTD for WD+RG and WD+MS systems from \citet{hachisu08}, the $\gamma-\alpha$ DD model of \citet{toonen12}, and a select wide DD DTD from \citet{greggio05}, the DTD from \citet{ruiter13} for violent mergers \citep{pakmor12} in the DD scenario \citep[similar to that of][but with updated cuts]{ruiter11}, and the double-detonation (SD) DTD from \citet{ruiter14}.

% FIGURE

Each of these studies predict multiple DTD outcomes achieved by varying physical assumptions about the physics of binary evolution, and thus produce subtle differences in the final shape of the \snia\ DTD \citep[for a thorough discussion of this topic, see the PopCORN project analysis in][]{popcorn}. The qualitative behavior of the final \snia\ age distribution follows the trends identified in Section~\ref{subsec:snia_age_dist}. In the young progenitor regime, the \snia\ age distribution follows the shape of the DTD. In the old progenitor regime, the age distribution shows a characteristic bump (the tardy component) of \sneia\ whose progenitor stars formed at the epoch of peak cosmic star formation. A transition of \snia\ ages with host mass is produced for all literature DTDs, even for the most unique DTD (the double-detonation scenario, which is not believed to comprise the \emph{entire} population of \sneia).  Thus we find the qualitative nature of our results are quite robust against any changes to the specific form of the \snia\ DTD.

%%%%%%%%%%%%%%%%%%%%%%%%%%%%%%%%%%%%%%%%%%%%%%%%%%%%%%%%%%%%%%%%%%%%%%%%%%%%%
\section{Conclusions}
\label{sec:conclusions}
%---------------------------
% new deeper insight into prompt and tardy SN Ia populations
Using observationally-informed empirical models for galaxy mass assembly, we glean new insight into the origin of the apparent bimodality of \snia\ ages. The ``prompt'' and ``tardy'' two-component (or ``A+B'') model arises from the bimodal age distribution of \sneia\ realized in nature as a consequence of galaxy star formation histories. We show that this bimodality persists to intermediate redshift ($z\sim0.5$) and make predictions about the evolution of the two components.

%(1) SF hosts: 
% - DTD is steeper than galaxy SFHs
% - SNe Ia ages peak at DTD peak
% - true at all epochs through cosmic history!!!
Prompt (young)  \sneia\ arise from actively star-forming galaxies, whose star formation histories have evolved slowly compared to the sharp $t^{-1}$ \snia\ delay time distribution. This results in star-forming galaxies producing \sneia\ predominantly from progenitors whose ages correspond to the peak of the \snia\ DTD, which holds true for all star-forming galaxies at all epochs of cosmic history. Thus prompt \sneia\ originate from similar progenitor ages in all star-forming galaxies at all redshifts, making them the most uniform subset of \sneia\ in the Universe.

%(2) passive hosts:
% - shut off SF in the distant past
% - here SF beats the DTD
% - SNe Ia ages are the past main SF time
% - presents different mean age at different redshifts!!!
% - doesn't become small fraction of population as expected!
Tardy (old) \sneia\ occur in galaxies whose star formation ceased in the distant past. It is only in environments lacking young stars where the \snia\ DTD does not dominate the shape of the \snia\ age distribution. Instead, the \snia\ ages correspond to the past epoch where all the galaxy's stars were formed. This past epoch of star formation is strongly dependent on the redshift being probed, meaning tardy \sneia\ originate from different progenitor age groups at different redshifts.

% Hubble residual implications!
These two galaxy SFH regimes which produce the prompt and tardy \sneia\ correspond to different galaxy mass scales. Typically, low-mass galaxies are actively star-forming and thus produce prompt \sneia, while massive galaxies have ceased star formation and thus produce tardy \sneia. The mean age of \sneia\ undergoes a sharp transition with host mass, similar to that observed in \snia\ Hubble residuals. If the observed Hubble residual step is indeed driven by progenitor age differences, then its magnitude should evolve in redshift in a manner which is not currently accounted for in \snia\ cosmology analyses.

% qualitative results are insensitive to the details!
These key results are qualitatively robust, and exhibit negligible sensitivity to the quantitative details of galaxy mass assembly or the \snia\ DTD. Future quantitative refinements will surely be possible with a better measurement of high redshift galaxy populations, particularly the fraction of passive galaxies as a function of stellar mass at very high redshifts. A precise measurement of the \snia\ delay time distribution would then enable quantitative predictions for the potential biases introduced in \snia\ cosmology analyses by the evolution of the progenitor age distribution.

%%%%%%%%%%%%%%%%%%%%%%%%%%%%%%%%%%%%%%%%%%%%%%%%%%%%%%%%%%%%%%%%%%%%%%%%%%%%%
\vskip11pt
% Acknowledgments
%\scriptsize
{\em Acknowledgments:}
We are very grateful to Richard Scalzo, Ashley Ruiter, Brian Schmidt, Fuyan Bian, Lee Spitler, Edward (Ned) Taylor, and Aaron Robotham for fruitful discussions. We thank Ashley Ruiter, Nikki Mennekens, Silvia Toonen, and Izumi Hachisu for providing digital representations of their \snia\ DTDs, Jonas Johansson for providing his \snia\ age-stretch relation, and Mark Sullivan and Yen-Chen Pan for providing the PTF host galaxy mass sample.
% referee!
We also thank the anonymous referee for thoughtful feedback on the text, and some very insightful suggestions which particularly strengthened the connection between our models and observations.
% funding (me + individuals)
% CAASTRO
This research was conducted by the Australian Research Council Centre of Excellence for All-sky Astrophysics (CAASTRO), through project number CE110001020.
% ADS
This research has made use of NASA's Astrophysics Data System (ADS).

%%%%%%%%%%%%%%%%%%%%%%%%%%%%%%%%%%%%%%%%%%%%%%%%%%%%%%%%%%%%%%%%%%%%%%%%%%%%%
% BIBLIOGRAPHY
%\bibliographystyle{mn2e}
\bibliographystyle{apj}
\bibliography{snia_age_vs_mass}

%%%%%%%%%%%%%%%%%%%%%%%%%%%%%%%%%%%%%%%%%%%%%%%%%%%%%%%%%%%%%%%%%%%%%%%%%%%%%
% APPENDIX ON GALAXY SCALING RELATION EQUATIONS
\appendix

%---------------------------
\section{Galaxy Mass Assembly: Parametrization of Models}
\label{app:equations}
%---------------------------
This Appendix presents the formal parametrizations for all input scaling relations employed in our galaxy mass assembly models.

%---------------------------
\subsection{Nominal SFH Parametrizations}
%---------------------------
% SMz relation...
The nominal relationship between stellar mass and star formation rate as a function of redshift (the SMz relation) is formally defined in Z12 as:
\begin{equation}
  \Psi(M_*, z) = 2.00\cdot\exp(1.33z)
  \left(\frac{M_*}{10^{10}}\right)^{0.7}
  [M_\odot\mathrm{yr}^{-1}]
\label{eq:smz}
\end{equation}

% mass loss (LK11)
Stellar mass loss as a function of time after epoch of star formation is given by LK11 for a variety of IMFs. For our \citet{chab03} IMF this parametrization is:
\begin{equation}
  f_{ml}(t) = 0.046 \ln\left(\frac{t}{0.276~\mathrm{Myr}}+1\right)
\label{eq:mass_loss}
\end{equation}

% Quenching vs. z
To accurately account for the redshift evolution of the quenching mass, we use the parametrization of \citet{muzzin13}, which is anchored at \logm$=10.55$ at $z=0.35$. Explicitly, our parametrization for the quenching mass as a function of redshift is:
\begin{equation}
  \log\left(\frac{M_Q(z)}{M_\odot}\right) = \left\{
    \begin{array}{rr}
       10.43+0.9\log(1+z) & : z \leq 1.5 \\
       8.56+5.6\log(1+z) & : z >  1.5 
    \end{array}
    \right.
\label{eq:mq_vs_z}
\end{equation}

\begin{figure}
\begin{center}
\includegraphics[width=0.45\textwidth]{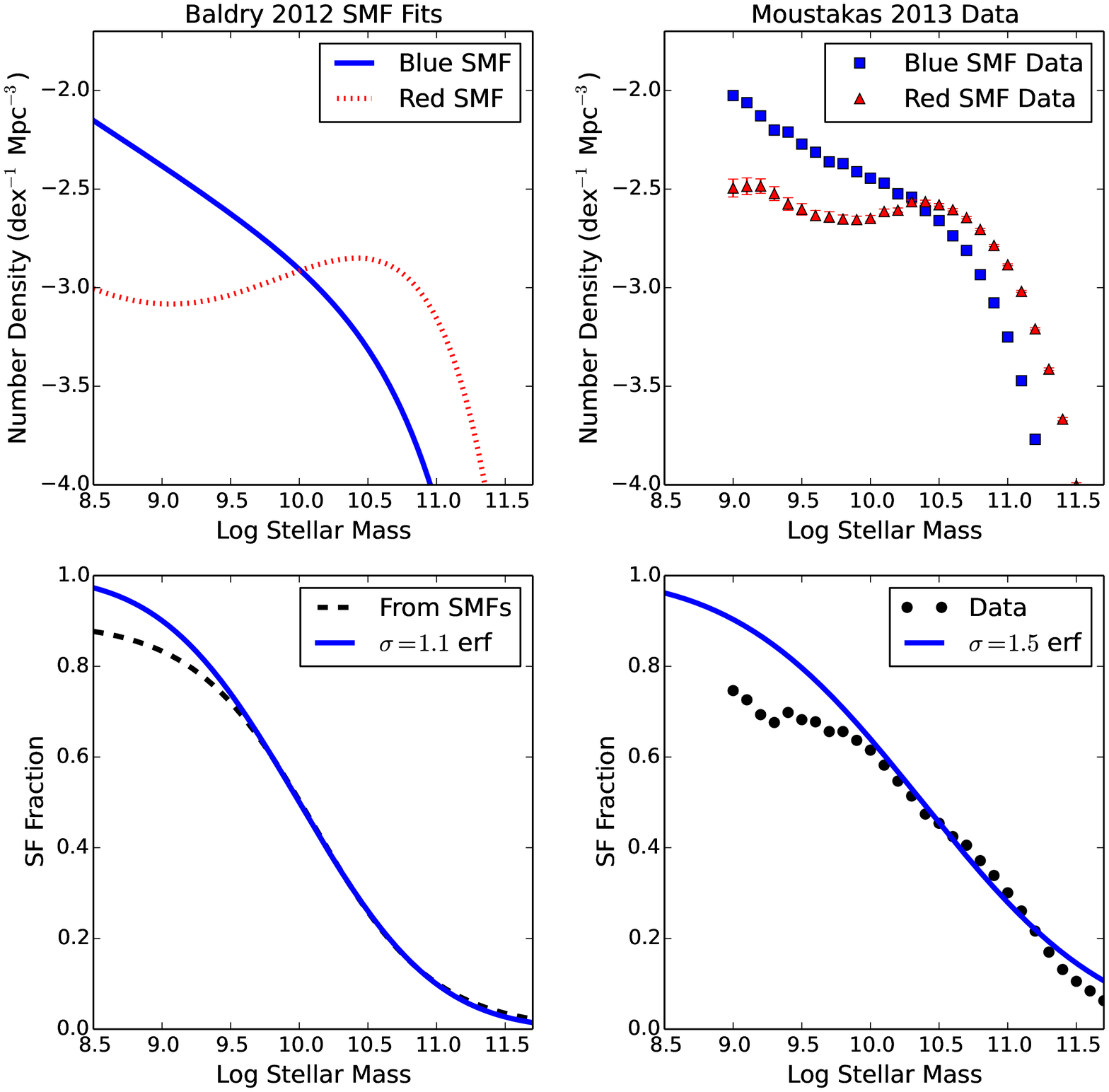}
\caption{Top left: blue and red galaxy stellar mass functions (SMFs) from GAMA \citep{baldry12}. Bottom left: blue (i.e., SF) galaxy fraction as a function of mass for GAMA (dashed black curve) and $erf$ model with width $\sigma_Q=1.1$. Top right: blue and red galaxy number densities as a function of mass for SDSS+{\it GALEX} from \citet{moustakas13}. Bottom right: blue (SF) galaxy fraction versus mass from data (black points) and $erf$ model with width $\sigma_Q=1.5$.}
\label{fig:quenching_penalty}
\end{center}
\end{figure}

% quenching vs. mass!!
To account for the gradual transition from active galaxies at low stellar mass to passive galaxies at high stellar mass, we impose a quenching penalty function of the form:
\begin{equation}
  p_Q(M_*, z) = \frac{1}{2}\left[1-erf\left(\frac{\log(M_*)-\log(M_Q(z))}{\sigma_Q}\right)\right]
\label{eq:quenching_penalty}
\end{equation}
where $M_Q(z)$ is the quenching mass as a function of redshift. Our nominal value for the quenching mass transition scale is $\sigma_Q=1.5$, which is a good representation of the active galaxy fraction for the galaxy stellar mass functions calculated in the low redshift universe by \citet{moustakas13}. In Figure~\ref{fig:quenching_penalty} we show these mass functions and the nominal $\sigma_Q=1.5$ quenching penalty function (right panels), as well as the ``narrow quenching'' width ($\sigma_Q=1.1$) penalty function which is a good fit to the GAMA \citep{baldry12} mass functions (left panels).

% FIGURE

% final mass buildup integration equation!
Finally, with all these parametrizations in hand (Equations~\ref{eq:smz}, \ref{eq:mass_loss}, \ref{eq:mq_vs_z}, \ref{eq:quenching_penalty}), we have the components necessary to explicitly integrate the galaxy stellar mass assembly using the following equation:
\begin{equation}
  \frac{M_*(t+\Delta t)-M_*(t)}{\Delta t} = 
   p_Q(M_*(t), z(t))\cdot\Psi(M_*(t), z(t)) - \frac{\Delta M_*}{\Delta t}
\label{eq:mass_buildup}
\end{equation}
where the mass lost in each time interval is a sum of mass lost from stars formed in each previous time step:
\begin{equation}
   \Delta M_* = \int_{0}^{t-t_f} \Psi(M_*(t-\tau), z(t-\tau))\cdot(f_{ml}(\tau+\Delta t)-f_{ml}(\tau)) d\tau
\end{equation}

%---------------------------
\subsection{Alternate SFH Parametrizations}
\begin{figure}
\begin{center}
\includegraphics[width=0.45\textwidth]{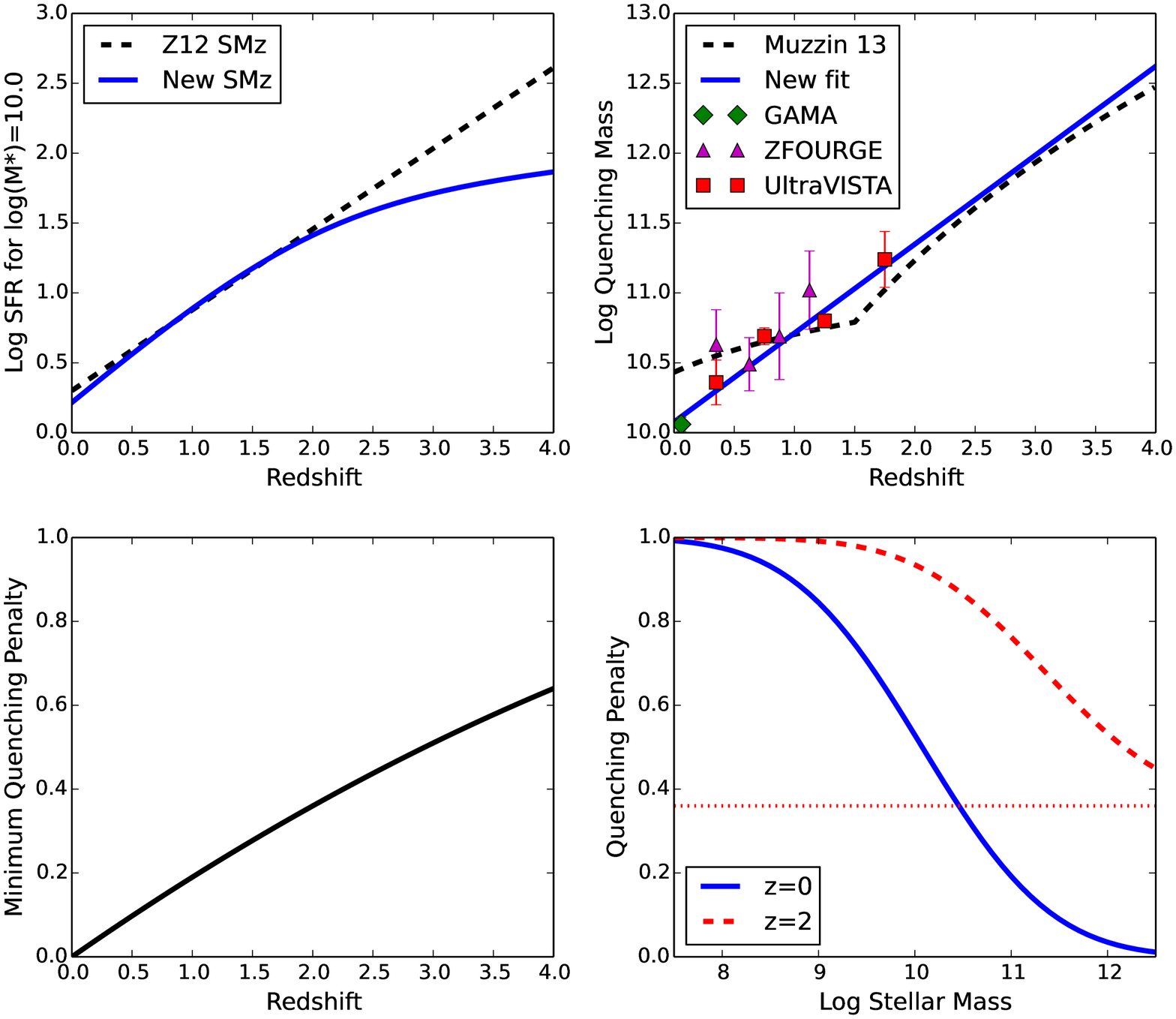}
\caption{Top left: SFR for a \logm\ $=10$ galaxy as a function of redshift for the Z12 SMz relation (dashed black line, Eq.~\ref{eq:smz}) and the modified SMz (solid blue line, Eq.~\ref{eq:alt_smz}). Top right: Quenching mass data (points), original quenching versus redshift formula (dashed black line, Eq.~\ref{eq:mq_vs_z}) from \citet{muzzin13}, and new linear fit to the data (solid blue line, Eq.~\ref{eq:alt_mq_vs_z}). Bottom left: Minimum quenching penalty versus redshift in for the ``fine-tuned'' model. Bottom right: quenching penalty versus mass at two redshifts (solid blue curve $z=0$, dashed red curve $z=2$) for the ``fine-tuned'' model (Eq.~\ref{eq:qpmin}).}
\label{fig:alt_sfh_eqs}
\end{center}
\end{figure}

%---------------------------
The alternate parametrizations for galaxy mass assembly model inputs are displayed graphically in Figure~\ref{fig:alt_sfh_eqs}, and are described in turn below.

% FIGURE

% adjusted SFR vs. z
Our alternate parametrization for the SMz relation is:
\begin{equation}
  \Psi(M_*, z) = 36.4\cdot
  \left(\frac{M_*}{10^{10}}\right)^{0.7}
  \cdot\frac{\exp(1.9z)}{\exp(1.7z)+\exp(0.2z)}
\label{eq:alt_smz}
\end{equation}
This is a hand-constructed formula which has the same mass dependence as the Z12 relation but a different redshift dependence. It matches the Z12 SMz relation well below $z=2$ but becomes much shallower above $z=2$.  In the top left panel of Figure~\ref{fig:alt_sfh_eqs}, we show the SFR for a $M_*=10^{10}M_\odot$ galaxy as a function of redshift for both parametrizations.

% re-fit quenching mass (give function)
Next we fit for a different functional form of the quenching mass as a function of redshift $M_Q(z)$ using several data sets. To do so, we use data sets where the blue and red stellar mass functions are well-fit and show a clear quenching mass scale where the quiescent fraction crosses 50\%. The final data sets we employ are the $z=0.06$ GAMA point \citep{baldry12}, the three lowest redshift bins from UltraVISTA \citep{muzzin13}, and the four lowest redshift points from ZFOURGE \citep{tomczak14}. These data points are extremely well fit by a linear function of the form:
\begin{equation}
  \log(M_Q(z)/M_\odot) = 10.077 + 0.636\cdot z
\label{eq:alt_mq_vs_z}
\end{equation}
The top right panel of Figure~\ref{fig:alt_sfh_eqs} shows the data points employed in our fit, the new fit $M_Q(z)$, and the original \citet{muzzin13} parametrization.

% nonzero minimum qp!
Finally, for our ``fine-tuned'' model we employed a minimum quenching penalty value which evolves quadratically from 0\% at redshift $z=0$ to 100\% at redshift $z=10$. Formally this manifests in the quenching penalty equations as:
\begin{eqnarray}
  \bar p_Q(M_*, z) & = & p_\mathrm{min}(z) + (1-p_\mathrm{min}(z))\cdot p_Q(M_*,z)
  \nonumber \\
  p_\mathrm{min}(z) & = & 1 - \left(\frac{z-10}{10}\right)^2
\label{eq:qpmin}
\end{eqnarray}
The bottom left panel of Figure~\ref{fig:alt_sfh_eqs} shows the minimum quenching penalty as a function of redshift, while the bottom right panel shows an example of the quenching penalty functions at $z=0$ and $z=2$ in the new parametrization.

%%%%%%%%%%%%%%%%%%%%%%%%%%%%%%%%%%%%%%%%%%%%%%%%%%%%%%%%%%%%%%%%%%%%%%%%%%%%%
\end{document}